\documentclass{article}
\usepackage{spconf,epsfig}
\usepackage{amssymb,amsmath,amscd,amsfonts,amstext}

\newcommand{\CC}{{\mathbb{C}}}
\newcommand{\EE}{{\mathbb{E}}}
\newcommand{\T}{{\mathrm{T}}}
\newcommand{\tr}{{\mathrm{tr}}}
\newcommand{\as}{\ \ {\text{a.s.}}}
\newcommand{\bs}{\boldsymbol}
\newcommand{\diag}{{\mathrm{diag}}}
\newcommand{\eqdef}{{\stackrel{\mathrm{def}}{=}}} 
\newtheorem{corollary}{Corollary}
\newtheorem{proposition}{Proposition}
\newtheorem{theorem}{Theorem}

\newtheorem{remark}{Remark}
\def\adots{
  \mathinner{\mkern1mu\raise1pt\hbox{.}\mkern2mu\raise4pt\hbox{.}
  \mkern2mu\raise7pt\vbox{\kern7pt\hbox{.}}\mkern1mu}}
\def\build#1_#2^#3{\mathrel{
\mathop{\kern 0pt#1}\limits_{#2}^{#3}}}
\title{A Central Limit Theorem for the SNR at the Wiener Filter Output for
Large Dimensional Signals} 
\name{Abla Kammoun$^{(1)}$, Malika Kharouf$^{(2)}$, 
Walid Hachem$^{(3)}$ and Jamal Najim$^{(3)}$\thanks{Work partially funded by the French 
`` Fonds National de la Science '', ACI/NIM project number 205 (MALCOM).}}
\address{$^{(1)}$ ENST (France), $^{(2)}$Casablanca University (Morocco), 
$^{(3)}$ CNRS/ ENST (France). \\
\texttt{(abla.kammoun, malika.kharouf, walid.hachem, jamal.najim@enst.fr)}}
\begin{document}
\ninept
\maketitle
\begin{abstract}
Consider the quadratic form 
$\beta = {\bf y}^* ( {\bf YY}^* + \rho {\bf I})^{-1} {\bf y}$ 
where $\rho$ is a positive number, where 
${\bf y}$ is a random vector and ${\bf Y}$ is a $N \times K$ random matrix both 
having independent elements with different variances, and where ${\bf y}$ and 
${\bf Y}$ are independent. Such quadratic forms represent the Signal to Noise 
Ratio at the output of the linear Wiener receiver for multi dimensional
signals frequently encountered in wireless communications and in array processing. 
Using well known results of Random Matrix Theory, the quadratic form $\beta$ 
can be approximated with a known deterministic real number $\bar\beta_K$
in the asymptotic regime where $K\to\infty$ and $K/N \to \alpha > 0$. This paper
addresses the problem of convergence of $\beta$. More specifically, it is shown 
here that $\sqrt{K}(\beta - \bar\beta_K)$ behaves for large $K$ like a Gaussian random
variable which variance is provided. 
\end{abstract}
\section{Introduction}
\label{sec:intro}
Consider the $N$ dimensional received signal 
$$
{\bf r} = {\bs \Sigma}{\bf s} + {\bf n} 
$$
where ${\bf s} = [s_0,s_1,\ldots,s_K]^\T$ is the transmitted complex
vector signal with size $K+1$ satisfying $\EE {\bf ss}^* = {\bf
  I}_{K+1}$, matrix ${\bs \Sigma}$ represents the ``channel'' in the
wide sense and ${\bf n}$ is the independent AWGN with covariance
matrix $\EE {\bf nn}^* = \rho {\bf I}_N > {\bf 0}$.  In this article,
we are interested in the performance of the linear Wiener estimate
(also called LMMSE for Linear Minimum Mean Squared Error estimate) of
signal $s_0$.  Among the various performance indexes, we shall focus
on the Signal to Noise Ratio (SNR) which can be expressed as follows:
Partition the channel matrix as ${\bs \Sigma} = \left[
  {\bf y} \ {\bf Y} \right]$, then the Wiener estimate $\hat s_0$ of
$s_0$ writes $\hat s_0 = {\bf y}^* \left( {\bs \Sigma}{\bs \Sigma}^* + \rho
  {\bf I}_N \right)^{-1} {\bf r}$ and the associated SNR $\beta_K$ is
given by:
$$
\beta_K = {\bf y}^* \left( {\bf YY}^* + \rho {\bf I}_N \right)^{-1} {\bf y}\ .
$$
A popular tool to address this problem, widely used in
multidimensional signal processing and communication engineering, is
represented by the theory of 
Large Random Matrices: Assume that ${\bs \Sigma}$ is random (in this
case, $\beta_K$ becomes a conditional SNR) and let $N\to\infty$ with
$K/N \to \alpha > 0$ (denoted in the sequel by ``$K\to\infty$'' for
short). As amply shown in the literature, there are many statistical
models related to ${\bs \Sigma}$ for which there exists a
deterministic sequence $\bar{\beta}_K$ such that $\beta_K -
\bar{\beta}_K \to 0$ almost surely (a.s.); this approximation is
generally defined as the solution of an implicit equation. Beyond the
convergence of the SNR, a natural practical and theoretical problem concerns
the study of its fluctuations (think for instance to the outage
probability evaluations). Despite its interest, there are very few related articles in the literature.
In this paper, we provide a Central Limit Theorem (CLT)
for $\beta_K$ as $K \to \infty$ for a general model of matrix ${\bs
  \Sigma}$: Assume that the $N \times (K+1)$ matrix ${\bs \Sigma}$ is
given by:
\begin{equation}
\label{eq-modele-sigma} 
{\bs \Sigma} = \frac{1}{\sqrt{K}} \left[ \sigma_{nk} W_{nk} \right]_{n=1,k=0}^{N,K} 
\end{equation} 
where $(\sigma_{nk}^2;\ 1\le n \le N;\ 0\le k \ldots,K)$ is a sequence of real 
numbers called a variance profile and where the complex random variables $W_{nk}$ are 
independent and identically distributed (i.i.d.) with $\EE W_{nk} = 0$, 
$\EE W_{nk}^2 = 0$, and $\EE | W_{nk} |^2 = 1$. 
In this case, the quadratic form $\beta_K$ is given by:
\begin{equation}
\label{eq-fq} 
\beta_K = \frac 1K 
{\bf w}_0^* {\bf D}_0^{1/2} 
\left( {\bf YY}^* + \rho {\bf I}_N \right)^{-1} {\bf D}_0^{1/2} {\bf w}_0 
\end{equation} 
where ${\bf w}_0 = [ W_{10}, W_{20}, \ldots, W_{N0} ]^\T$
and ${\bf D}_0$ is the $N\times N$ diagonal nonnegative matrix ${\bf D}_0 = 
\diag(\sigma^2_{10},\ldots, \sigma^2_{N0})$. 
An important special case that we shall describe carefully in the sequel is when
the variance profile is {\em separable}, i.e $\sigma^2_{nk} = d_n \tilde d_k$. \\

Among the many applications of the general model \eqref{eq-modele-sigma}, let us mention:
\begin{itemize}
\item 
Multiple antenna transmissions with $K+1$ antennas at the transmission side 
and $N$ antennas at the reception side. Here we consider the transmission model
${\bf r} = {\bs \Xi}{\bf s} + {\bf n}$ where 
${\bs \Xi} = \frac{1}{\sqrt{K}}{\bf H} {\bf P}^{1/2}$, matrix ${\bf H}$ is a $N \times (K+1)$ 
random matrix with complex Gaussian elements representing the radio channel and 
${\bf P} = \diag(p_0, \ldots, p_K)$ is the (deterministic) matrix of the powers given to the
different users. Write ${\bf H} = \left[ {\bf h}_0 \ \cdots \ {\bf h}_K \right]$, and
assume the columns ${\bf h}_k$ are independent, which is realistic
when the transmitters are distant one from another. Let ${\bf C}_k$ be the covariance
matrix ${\bf C}_k = \EE {\bf h}_k {\bf h}_k^*$ and let ${\bf C}_k = {\bf U}_k 
{\bs \Lambda}_k {\bf U}_k$ be a spectral decomposition of ${\bf C}_k$ 
where ${\bs \Lambda}_k = \diag((\lambda_{nk})_{n=1,\ldots,N})$ is the matrix of
eigenvalues. 
If the eigenvector matrices ${\bf U}_0, \ldots, {\bf U}_K$ are all equal 
(note that sometimes they are all identified with the Fourier $N \times N$ 
matrix \cite{say-sp02}),
then one can show that matrix ${\bs \Xi}$ introduced above can be replaced with
matrix ${\bs \Sigma}$ of Model \eqref{eq-modele-sigma} where the $W_{nk}$ are standard
Gaussian iid and $\sigma^2_{nk} = \lambda_{nk} p_k$. \\ 
In certain situations it is furthermore assumed that 
${\bs \Lambda}_0 = \cdots = {\bs \Lambda}_K = 
\diag(\lambda_1, \ldots, \lambda_N)$: this is the well known Kronecker model
with correlations at reception. In our setting this model is accounted for 
by the separable variance profile case $\sigma^2_{nk} = \lambda_{n} p_k$.

\item CDMA transmissions on flat fading channels. 
Here $N$ is the spreading factor, $K+1$ is the number of users, and  
\begin{equation}
\label{eq-cdma-flat} 
{\bs \Sigma} = {\bf W} {\bf P}^{1/2} 
\end{equation} 
where ${\bf W}$ is the $N \times (K+1)$ signature matrix assumed here to have
random i.i.d. elements with mean zero, variance $1/N$ and 
where ${\bf P} = \diag(p_0, \ldots, p_K)$ is the users powers matrix. 
In this case, the variance profile is separable with $d_n = 1$ and 
$\tilde d_k = \frac KN p_k$. 

\item 
MC-CDMA transmissions on frequency selective channels.
In the uplink, the matrix ${\bs \Sigma}$ is written:
$$
{\bs \Sigma} = \left[ {\bf H}_0 {\bf w}_0 \ \cdots \ 
{\bf H}_{K+1} {\bf w}_{K+1} \right] 
$$
where ${\bf H}_k = \diag(h_k(\exp(2 \imath \pi (n-1) /
N))_{n=1,\ldots,N})$ is the radio channel matrix of user $k$ in the
discrete Fourier domain and ${\bf W} = [ {\bf w}_1, \cdots, {\bf w}_K
]$ is the $N \times (K+1)$ signature matrix with iid elements 
as in the CDMA case above. Modeling this time the channels
transfer functions as deterministic functions, we have $\sigma^2_{nk}
= \frac KN | h_k(\exp(2 \imath \pi (n-1) / N)) |^2$. \\
In the downlink, we have
\begin{equation}
\label{eq-mccdma-downlink} 
{\bs \Sigma} = {\bf H} {\bf W} {\bf P}^{1/2} 
\end{equation}
where ${\bf H} = \diag(h(\exp(2 \imath \pi (n-1) / N))_{n=1,\ldots,N})$ is the
radio channel matrix in the discrete Fourier domain, the $N \times (K+1)$ 
signature matrix ${\bf W}$ is as above, and 
${\bf P} = \diag(p_0, \ldots, p_K)$ is the matrix of the powers given to the
different users. Model \eqref{eq-mccdma-downlink} coincides with the separable
variance profile case with 
$d_n = \frac KN | h(\exp(2 \imath \pi (n-1) / N)) |^2$ and $d_k = p_k$. 
\end{itemize} 

\paragraph*{About the Literature.} 
In the communication engineering literature, the CLT for the quadratic forms
has been considered probably for the first time in \cite{tse-zei-it00}, where the 
authors consider the case where ${\bs \Sigma}$ is a matrix with
i.i.d. elements. Their results are based on \cite{sil-ap90} where
the asymptotic behaviour of the eigenvectors 
of ${\bs \Sigma} {\bs \Sigma}^*$ is described.  
Recently, \cite{pan-guo-zhou-aap07} considered the more general CDMA Model 
\eqref{eq-cdma-flat}. 
The model considered in this paper includes the models of \cite{tse-zei-it00} and 
\cite{pan-guo-zhou-aap07} as special cases. 
The approach used here to establish the CLT is powerful yet simple. 
It is based on the representation of $\beta_K$ as the sum of a martingale
difference sequence and the use of the CLT for martingales \cite{bil-PM-livre95}. 
\paragraph*{} 

This paper is organized as follows. 
In Section \ref{sec-1st-order} we recall the first order results that describe
the limiting behaviour of $\beta_K$. The CLT for $\beta_K$ is stated in Section \ref{sec-clt}.
A sketch of proof for the CLT is presented in Section \ref{sec-proof}.
Finally, we provide simulations in Section \ref{sec-simus}. 

\section{SNR Deterministic Approximation}
\label{sec-1st-order}

Let us begin with a definition and some notations. 
We say that a complex function $t(z)$ belongs to class
${\cal S}$ if $t(z)$ is analytical in the upper half plane 
$\CC_+ = \{ z \in \CC \ : \ \text{im}(z) > 0\}$, if $t(z) \in \CC_+$ for
all $z \in \CC_+$ and if $\text{im}(z) | t(z) |$ is bounded over 
$\CC_+$. 
We introduce the diagonal matrices 
\begin{eqnarray*} 
{\bf D}_k &=& \diag(\sigma^2_{1k}, \ldots, \sigma^2_{Nk}),  \quad  
k=1,\ldots, K \\
\widetilde{\bf D}_n &=& \diag(\sigma^2_{n1}, \ldots, \sigma^2_{nK}),  \quad  
n=1,\ldots, N  
\end{eqnarray*} 
and the diagonal matrix functions
\begin{eqnarray*} 
{\bf T}(z) &=& \diag(t_1(z), \ldots, t_N(z)) \\ 
\widetilde{\bf T}(z) &=& \diag(\tilde t_1(z), \ldots, \tilde t_K(z))
\end{eqnarray*} 
that are specified by the following proposition: 
\begin{proposition}
(\cite{gir-livre90,hac-lou-naj-aap07}) 
\label{prop-def-T-tilde(T)} 
The system of $N+K$ functional equations
$$
\left\{\begin{array}{lcl} 
t_n(z) &=& \displaystyle{\frac{-1}
{z\left( 1 + \frac 1K \tr (\widetilde{\bf D}_n \widetilde{\bf T}(z))\right)}, 
\quad 1\leq n \leq N} \\
\tilde{t}_k(z) &=& \displaystyle{\frac{-1}
{z\left( 1 + \frac 1K \tr ({\bf D}_k {\bf T}(z))\right)}, 
\quad 1\leq k \leq K}  
\end{array} \right. 
$$
has a unique solution $({\bf T}, \widetilde{\bf T})$ among the 
diagonal matrices for which the $t_n$ and the $\tilde t_k$ belong to class
${\cal S}$. 
Functions $t_n(z)$ and $\tilde{t}_k(z)$ so defined admit analytical 
continuations over $\CC - [0, \infty)$.
\end{proposition}
In the separable case, we have ${\bf D}_k = \tilde d_k {\bf D}$ and 
$\widetilde{\bf D}_n = d_n \widetilde{\bf D}$ where ${\bf D} = \diag(d_1, \ldots, d_N)$ and
$\widetilde{\bf D} = \diag(\tilde d_1, \ldots, \tilde d_K)$. In this case, the system
described above simplifies to a system of two equations: 
\begin{proposition}
\label{prop-first-order-separable} 
The system of two functional equations 
\begin{equation} 
\label{eq-first-order-separable} 
\left\{\begin{array}{lcl} 
\delta(z) &=&  
\frac{1}{K} \tr \left( 
{\bf D} \left( -z({\bf I}_N + \tilde\delta(z) {\bf D}) \right)^{-1} \right)  \\
\tilde\delta(z) &=&  
\frac{1}{K} \tr \left( 
\widetilde{\bf D} \left( -z({\bf I}_K + \delta(z) \widetilde{\bf D}) \right)^{-1} \right)
\end{array}\right.  
\end{equation} 
admits a unique solution $(\delta, \tilde{\delta}) \in {\cal S}^2$. Moreover, letting
$z = -\rho \in (-\infty,0)$, the system admits a unique pointwise solution
$(\delta(-\rho), \tilde\delta(-\rho))$ such that 
$\delta(-\rho) > 0$, $\tilde\delta(-\rho) > 0$. 
\end{proposition} 
In this particular case, the matrix functions ${\bf T}$ and $\widetilde{\bf T}$ defined 
by Proposition \ref{prop-def-T-tilde(T)} are given by 
${\bf T} = -\frac 1z ({\bf I} + \tilde\delta {\bf D} )^{-1}$ and 
$\widetilde{\bf T} = -\frac 1z ({\bf I} + \delta \widetilde{\bf D} )^{-1}$.  
The asymptotic behaviour of $\beta_N$ is characterized by the following 
theorem: 
\begin{theorem}
\label{th-1st-order}
(\cite{gir-livre90, li-tul-ver-it04, hac-lou-naj-aap07}) 
Let $\bar\beta_K = \displaystyle{
\frac1K \tr {\bf D}_0 {\bf T}(-\rho)}$ where ${\bf T}$ is
given by Proposition \ref{prop-def-T-tilde(T)}. Then 
$$
\beta_K - \bar\beta_K 
\xrightarrow[K\to\infty]{} 0 \quad \text{almost surely.} 
$$
\end{theorem} 
\begin{remark}
In matrix model \eqref{eq-modele-sigma}, one sometimes assumes that the
variance profile $\sigma^2_{nk}$ is obtained from the samples of a continuous nonnegative
function $\pi(x,y)$ defined on $[0,1]^2$ at points $(n/N,k/(K+1))$, i.e.
$\sigma^2_{nk} = \pi(n/N, k/(K+1))$. In this particular case, the sequences $\bar\beta_K$ and
$\delta_K$ defined in Theorem \ref{th-1st-order} above (and also Corollary \ref{cor-1st-order-separable} below)
converge to limits that are solutions of integral equations
(see for instance \cite{li-tul-ver-it04,cha-hac-lou-it04}). 
\end{remark}

In the separable case, ${\bf D}_0 = \tilde d_0 {\bf D}$ hence we have
\begin{corollary}
\label{cor-1st-order-separable}
(\cite{li-tul-ver-it04,cha-hac-lou-it04}) 
Assume the separable case $\sigma^2_{nk} = d_n \tilde d_k$. Then
$$
\frac{\beta_K}{\tilde d_0} - \delta_K \xrightarrow[K\to\infty]{} 0 \quad \as 
$$
where $\delta_K = \delta$ with $(\delta, \tilde\delta)$ being the solution of 
System \eqref{eq-first-order-separable} at $z=-\rho$. 
\end{corollary}

\begin{remark} (see also Corollary \ref{clt-separable} below) In the
  separable case, $\beta_K / \tilde d_0$ often represents the SNR of
  user $0$ normalized to this user's power. Therefore,  
  we can naturally interpret the approximation $\delta_K$ as
  an asymptotic normalized SNR. This approximation, as well
  as the asymptotic variance of the normalized SNR $\beta_K / \tilde d_0$ 
  defined in Corollary \ref{clt-separable} is the same
  for all users. 
\end{remark}


\section{SNR Fluctuations: the CLT}  
\label{sec-clt}
We now come to the main result of this paper, which holds true under some slight 
technical assumptions: 
\begin{theorem}
\label{th-main} 
Let ${\bf A}$ and ${\bs \Delta}$ be the $K\times K$ matrices 
\begin{eqnarray*} 
{\bf A} &=& 
\left[ 
\frac 1K \frac{\frac{1}{K} \tr {\bf D}_{\ell} {\bf D}_{m} {\bf T}(-\rho)^2}
{\left(1 + \frac{1}{K} \tr {\bf D}_{\ell} {\bf T}(-\rho) \right)^2} 
\right]_{l,m=1}^K \quad \text{and} \\
{\bs \Delta} &=& 
\diag\left( \left( 1 + \frac 1K \tr {\bf D}_l {\bf T}(-\rho) \right)^2_{l=1,\ldots,K} 
\right) 
\end{eqnarray*} 
where ${\bf T}$ is defined by Proposition \ref{prop-def-T-tilde(T)}. 
Let ${\bf g}$ be the $K \times 1$ vector
$$
{\bf g} = \left[ \frac 1K \tr {\bf D}_0 {\bf D}_1 {\bf T}(-\rho)^2, \cdots, 
\frac 1K \tr {\bf D}_0 {\bf D}_K {\bf T}(-\rho)^2 \right]^\T 
$$
Then the following hold true : 
\begin{itemize}
\item[1)] 
The sequence of real numbers 
\begin{multline} 
\Theta_K^2 =
( \EE| W_{10} |^4 - 1 )
\frac 1K \tr {\bf D}_0^2 {\bf T}^2 \\ 
+ \frac{1}{K} {\bf g}^\T ( {\bf I}_K - {\bf A})^{-1} {\bs \Delta}^{-1} {\bf g} 
\label{eq-theta} 
\end{multline} 
is well defined and furthermore 
$$
0 < \liminf_K \Theta_K^2 \leq \limsup_K \Theta_K^2 < \infty 
$$
\item[2)] 
The sequence $\beta_K$ satisfies 
$$
\sqrt{K} \frac{\beta_K - \bar\beta_K}{\Theta_K} 
\xrightarrow[K\to\infty]{} {\cal N}(0,1) 
$$
in distribution where $\bar\beta_K$ is defined in the statement of Theorem \ref{th-1st-order}. 
\end{itemize}
\end{theorem}
In the separable case, one can show that $\Theta_K^2 = \tilde d_0^2 \Omega_K^2$ where
$\Omega_K^2$ is given by the following corollary: 
\begin{corollary}\label{clt-separable}
Assume the separable case $\sigma^2_{nk} = d_n \tilde d_k$. 
Let $\gamma = \frac 1K \tr {\bf D}^2 {\bf T}^2$ and $\tilde\gamma = \frac 1K 
\tr \widetilde{\bf D}^2 \widetilde{\bf T}^2$. 
The sequence 
$$
\Omega_K^2 = 
\gamma \left(
\left( \EE|W_{10} |^4 - 1 \right) +
\frac{\rho^2 \gamma \tilde\gamma}{ 1 - \rho^2 \gamma \tilde\gamma}
\right)
$$
satisfies $0 < \liminf_K \Omega_K^2 \leq \limsup_K \Omega_K^2 < \infty$, and
$$
\sqrt{K} \frac{\beta_K/\tilde d_0  - \delta_K}{\Omega_K} 
\xrightarrow[K\to\infty]{} {\cal N}(0,1) 
$$
in distribution.
\end{corollary} 
\begin{remark}
These results show in particular that the asymptotic variance $\Theta_K^2$ is minimum 
with respect to the distribution of the $W_{nk}$ when $|W_{nk}| = 1$
with probability one. In the context of CDMA and MC-CDMA, this will be the case
when the signature matrix elements have their values in a PSK constellation. 
\end{remark} 

\section{Sketch of proof}
\label{sec-proof}
Let ${\bf Q}$ be the $N\times N$ matrix 
${\bf Q} = \left( {\bf YY}^* + \rho {\bf I}_N \right)^{-1}$. 
Recall that the deterministic approximation of $\beta_K$ is $\bar\beta_K = 
\frac 1K \tr {\bf D}_0 {\bf T}$. 
Getting back to Equation \eqref{eq-fq}, we can write
\begin{eqnarray*}
\sqrt{K}( \beta_K - \bar\beta_K ) &=& 
\frac{1}{\sqrt{K}} \left( {\bf w}_0^* {\bf D}_0^{1/2} {\bf Q} {\bf D}_0^{1/2} {\bf w}_0 
- \tr {\bf D}_0 {\bf Q} \right) \\
& & + \frac{1}{\sqrt{K}} \tr {\bf D}_0 \left( {\bf Q} - {\bf T} \right) \\
&\eqdef& \xi_K + \chi_K 
\end{eqnarray*} 
It can be shown \cite{hac-lou-naj-(sub)aap07} that $\EE \chi_K^2 = {\cal O}(1/K)$. On the
other hand, by using the independence of ${\bf w}_0$ and ${\bf Q}$ and the fact that the elements of 
${\bf w}_0$ are i.i.d., one can easily show that
$\EE \xi_K^2 = {\cal O}(1)$ as $K\to\infty$. As a consequence,
the asymptotic behaviour of $\sqrt{K}( \beta_K - \bar\beta_K )$ is given by $\xi_K$. Denote by $\EE_n$ the
conditional expectation $\EE_n[.] = \EE[. \| W_{n,0}, W_{n+1,0}, \ldots, W_{N,0}, 
{\bf Y} ]$. Put $\EE_{N+1}[.] = \EE[.\| {\bf Y}]$ and note that 
$\EE_{N+1} {\bf w}_0^* {\bf D}_0^{1/2} {\bf Q} {\bf D}_0^{1/2} {\bf w}_0 = 
\tr {\bf D}_0 {\bf Q}$. With these notations at hand, we have 
$$
\xi_K = 
\sum_{n=1}^N (\EE_n - \EE_{n+1}) 
\frac{{\bf w}_0^* {\bf D}_0^{1/2} {\bf Q} {\bf D}_0^{1/2} {\bf w}_0}{\sqrt{K}}  
\eqdef \sum_{n=1}^N Z_n \ .
$$
The sequence $Z_n$ is readily a martingale difference sequence with respect
to the increasing sequence of $\sigma-$fields $\sigma({\bf Y}), 
\sigma(W_{N,0}, {\bf Y})),$ $\ldots, \sigma(W_{1,0},\ldots, W_{N,0}, {\bf Y})$. 
The asymptotic behaviour of $\xi_K$ (convergence in distribution toward a Gaussian r.v. and 
derivation of the variance $\Theta_K^2$) can be characterized with the help of the
CLT for martingales \cite[Ch. 35]{bil-PM-livre95}. 
\section{Simulations}
\label{sec-simus}

In this section, the accuracy of the Gaussian approximation is verified by simulation. 
We consider an MC-CDMA transmission in the uplink direction. 
The base station detects the symbols of a given user in the presence of $K$ interfering users. 
We assume that the  discrete channel impulse response of each user consists in  
$L=5$ iid Gaussian coefficients with variance $1/L$. All impulse responses are
known to the base station. \\ 
In this case, ${\bs \Sigma}$ is given by:
$$
{\bs \Sigma} = \left[ \sqrt{{p}_0}{\bf H}_0 {\bf w}_0 \ \cdots \ 
\sqrt{p_{K+1}}{\bf H}_{K+1} {\bf w}_{K+1} \right] 
$$ where 
\begin{itemize}
\item ${\bf H}_k = \diag(h_k(\exp(2 \imath \pi (n-1) /,
N))_{n=1,\ldots,N})$ is the channel matrix of user $k$ in the frequency domain,
\item   $p_k$ is the amount of  power allocated to user $k$,
\item ${\bs w}_k$  are assumed to belong to QPSK constellation with mean zero and variance 
$1/N$.
\end{itemize}
In this case, $\sigma_{n,k}^2$ is given by:
$$
\sigma_{n,k}^2=\frac{Kp_k}{N}|h_k\left(\exp\left(2i\pi(n-1)/N\right)\right)|^2
$$
We denote by $P$ the power given to the user of interest. The other users are arranged into $5$ classes according to their powers. The power of each class as well as the proportion of users
within this class are given in table \ref{tab:class}.

\begin{figure}[htbp]
   \begin{center}
      \includegraphics[scale=0.5]{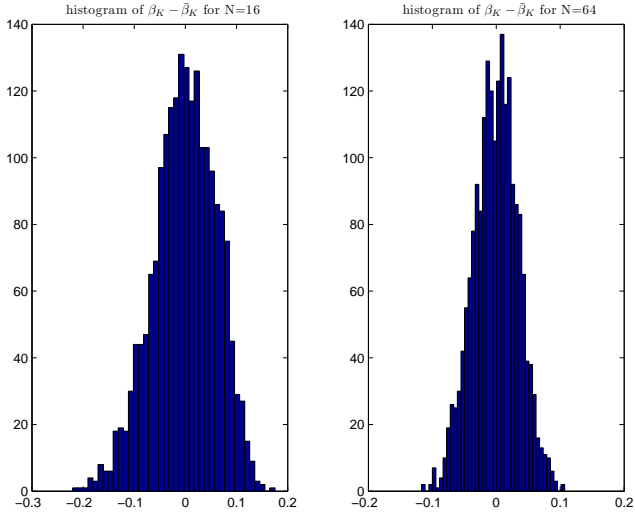}
   \end{center}
   \caption{\footnotesize Histogram of $\sqrt{K}(\beta_K-\bar{\beta}_K)$ 
   for $N=16$ and $N=64$.}
   \label{fig:histogram}
\end{figure}

\begin{figure}
 \begin{center}
      \includegraphics[scale=0.5]{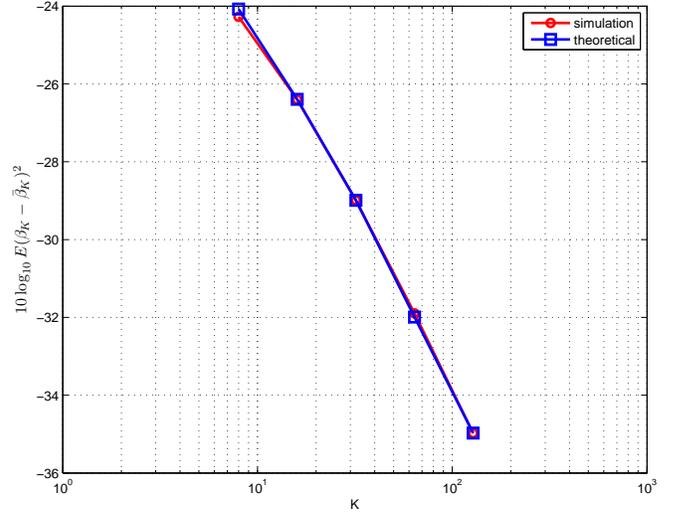}
   \end{center}
   \caption{\footnotesize Second moment of $\beta_K-\bar{\beta}_K$ }
   \label{fig:decroissance}

\end{figure}

\begin{table}[htbp]
\caption{Power and proportion of each user class}
\begin{center}
\begin{tabular}{|c|c|c|c|c|c|}
\hline
class & 1 &2 & 3 & 4 & 5 \\
\hline
Power & $P$ & $2P$ & $4P$ & $8P$ & $16P$ \\
\hline
Proportion & $\frac{1}{8}$ & $\frac{1}{4}$ & $\frac{1}{4}$ & $\frac{1}{8}$ & $\frac{1}{4}$\\
\hline
\end{tabular}
\end{center}
\label{tab:class}
\end{table}
Figure \ref{fig:histogram} shows the histogram of 
$\sqrt{K}(\beta_K-\bar{\beta}_K)$ for $N=16$ and $N=64$. 
We note that as it was predicted by our derived results, the histogram of 
$\sqrt{K}(\beta_K-\bar{\beta}_K)$ is similar to that of a Gaussian random variable. 
In Figure \ref{fig:decroissance} the measured second moment of 
$\beta_K-\bar{\beta}_K$ is compared with ${\Theta}_K^2 / K$. 
We note that convergence is reached even for $K=8$. 

\section{Conclusion}
The Gaussian character of the SNR at the output of the Wiener receiver
for a class of large dimensional signals described by a random 
transmission model has been established theoretically and verified
by simulation. 

\bibliographystyle{IEEEbib}
\bibliography{BSTLabrconf,bibli}

\def\cprime{$'$}
\begin{thebibliography}{10}

\bibitem{say-sp02}
A.M. Sayeed,
\newblock ``{Deconstructing Multiantenna Fading Channels},''
\newblock {\em IEEE Trans. on SP}, vol. 50, no. 10, pp. 2563--2579, Oct. 2002.

\bibitem{tse-zei-it00}
D.N.C. Tse and O.~Zeitouni,
\newblock ``{Linear Multiuser Receivers in Random Environments},''
\newblock {\em IEEE Trans. on IT}, vol. 46, no. 1, pp. 171--188, Jan. 2000.

\bibitem{sil-ap90}
J.W. Silverstein,
\newblock ``{Weak Convergence of Random Functions Defined by the Eigenvectors
  of Sample Covariance Matrices},''
\newblock {\em Ann. Probab.}, vol. 18, no. 3, pp. 1174--1194, 1990.

\bibitem{pan-guo-zhou-aap07}
G.-M. Pan, M.-H Guo, and W.~Zhou,
\newblock ``{Asymptotic Distributions of the Signal-to-Interference Ratios of
  LMMSE Detection in Multiuser Communications},''
\newblock {\em Ann. Appl. Probab.}, vol. 17, no. 1, pp. 181--206, 2007.

\bibitem{bil-PM-livre95}
P.~Billingsley,
\newblock {\em {Probability and Measure}},
\newblock John Wiley, 3rd edition, 1995.

\bibitem{gir-livre90}
V.~L. Girko,
\newblock {\em {Theory of Random Determinants}},
\newblock Kluwer, Dordrecht, 1990.

\bibitem{hac-lou-naj-aap07}
W.~Hachem, P.~Loubaton, and J.~Najim,
\newblock ``{Deterministic Equivalents for Certain Functionals of Large Random
  Matrices},''
\newblock {\em Ann. Appl. Probab.}, vol. 17, no. 3, pp. 875--930, 2007.

\bibitem{li-tul-ver-it04}
L.~Li, A.M. Tulino, and S.~Verd{\'u},
\newblock ``{Design of Reduced-Rank MMSE Multiuser Detectors Using Random
  Matrix Methods},''
\newblock {\em IEEE Trans. on IT}, vol. 50, no. 6, pp. 986--1008, June 2004.

\bibitem{cha-hac-lou-it04}
J.M. Chaufray, W.~Hachem, and Ph. Loubaton,
\newblock ``{Asymptotic Analysis of Optimum and Sub-Optimum CDMA Downlink MMSE
  Receivers},''
\newblock {\em IEEE Trans. on IT}, vol. 50, no. 11, pp. 2620--2638, Nov. 2004.

\bibitem{hac-lou-naj-(sub)aap07}
W.~Hachem, Ph. Loubaton, and J.~Najim,
\newblock ``{A CLT For Information-Theoretic Statistics of Gram Random Matrices
  with a Given Variance Profile},''
\newblock submitted to \emph{Ann. Appl. Probab.}, \texttt{arXiv:0706.0166}.

\end{thebibliography}

\end{document}